**Optical sound pressure measurement using Fabry-Pérot cavity for primary acoustic standards**

Koto Hirano,[1,a] Wataru Kokuyama,[1] Hajime Inaba,[1] Sho Okubo,[1] Tomofumi Shimoda,[1] Hironobu Takahashi,[1] Keisuke Yamada,[1] and Hideaki Nozato[1]

[1] *National Metrology Institute of Japan (NMIJ), National Institute of Advanced Industrial Science and Technology (AIST), Tsukuba, Ibaraki 305-8563, Japan*

Abstract

Optical sound pressure measurement is a promising technology to establish primary acoustic standards without reliance on specific types of microphones. We developed a precision optical sound pressure measurement system by combining a Fabry-Pérot optical cavity, a phase-stabilized optical frequency comb, and a custom-made phasemeter. The optical cavity detects changes in the air's refractive index induced by sound waves as changes in its resonance frequency. A continuous-wave laser frequency is stabilized at the resonance, and the frequency comb detects the changes in the laser frequency. The frequency changes are measured with high sensitivity and accuracy using a phasemeter that we developed. The sound pressures measured by this system agreed with the measurement value obtained using a reference microphone within 5% at sound pressure levels of 78 dB and 84 dB, within a frequency range of 100 Hz to 1 kHz. A systematic deviation of 2.6% was observed, with the optical system yielding higher values than the microphone. To identify the cause of this deviation, we performed vibration displacement measurements of the cavity mirrors and finite element analysis, which revealed that fluctuations in the optical path length due to insufficient fixation of the mirrors were responsible.

[a] Email: koto.hirano@aist.go.jp

# I. INTRODUCTION

In airborne acoustics, metrological standards have been established by supplying the microphone sensitivity [1], i.e., the ratio of the open-circuit voltage to sound pressure (V/Pa) rather than measuring the sound pressure directly. This is due to the difficulty associated with realizing a stable sound source with a constant sound pressure level over a wide frequency range, whereas microphone sensitivity is relatively stable over time. The traceable chain based on microphone sensitivity is operated as follows. National Metrology Institutes (NMIs) perform primary calibrations on specific microphones to assign their sensitivities, which then serve as reference values for comparison calibration of various acoustic measurements devices.

Although such frameworks have been well-established and operated for decades, there are several issues with microphone-dependent systems that must be considered. First, sensitivities belong to mechanical artifacts; thus, variations between individual microphones and long-term drift require periodic calibrations [2]. Second, the primary calibration of microphone sensitivity determines the sensitivity of specific microphone types, without directly measuring sound pressure. Current primary microphone calibration is based on reciprocity calibration methods, where the sensitivities of a set of three microphones are determined by measuring the ratio of the output voltage to the input current when one microphone is used as a receiver and another as a transmitter [3]. However, this approach lacks direct sound pressure measurement and is limited to stable, reciprocal microphones, such as laboratory standard microphones specified in IEC 61094-1:2000 [1].

Realization of sound pressure measurement using optical techniques addresses these limitations. First, this enables sound pressure (rather than artifact-associated sensitivity), to be treated as the primary measurand. Second, if this measurement system is applied to calibrate microphones, the dependence on specific, reciprocal microphones in the primary microphone calibration process can be eliminated. In addition, this approach can significantly broaden the scope of acoustic measurement applications because the optical system can function as a new type of sound measurement device. Thus, NMIs and research groups have been investigating optical sound pressure measurement techniques in the audible frequency range. One method is based on photon correlation techniques, where the velocity of micro-particles is measured and multiplied by the acoustic transfer impedance to derive the sound pressure [4, 5]. Since it is difficult to directly and accurately measure the velocity of gas molecules, these methods typically track particles with diameters of approximately 1 μm, thus relying on observing the motion of physical artifacts.

Another method involves detecting variations in the refractive index of air caused by pressure fluctuations via the acousto-optic effect [6]. While this technique enables the non-contact detection of sound, the change in the refractive index induced by sound is extremely small, on the order of $10^{-9}$/Pa, which requires special care to ensure a sufficient signal-to-noise ratio [7]. To overcome these challenges, a recently proposed method utilizes a Fabry-Pérot optical cavity to amplify the phase shift of light caused by refractive index changes, and then measures the resulting cavity resonance frequency shift. This method offers the significant advantage of utilizing frequency—the physical quantity with the lowest measurement uncertainty— measurement; however, it remains at the proof-of-concept stage. For example, Chijioke et al. reported that their cavity structure can only be applied to frequencies of $1 \times N$ kHz (where $N$ is an integer) [8], and Ishikawa et al. observed relatively large deviations (up to 30%) from the sound pressure levels measured by a microphone in the 100 Hz to 1 kHz frequency range [9].

In this study, two innovations were designed and developed: a mechanically robust Fabry-Pérot cavity and a frequency measurement system consisting of an optical frequency comb and a phasemeter. The mechanically robust cavity contributed to isolating only the refractive index change induced by sound, while the phasemeter equipped with a proprietary algorithm enabled precise measurement of the rapid and large-amplitude phase evolution. Together, this measurement method enabled more precise sound pressure measurements over a wider frequency range than in previous studies [8, 9]. We measured sound pressure of 78 dB or 84 dB (ref. 20 µPa) from 100 Hz to 5 kHz. In the 100 Hz to 1 kHz range, the sound pressures obtained using the proposed system agreed with those obtained using a reference microphone within a deviation of 5%. This deviation was primarily attributed to the change in the cavity's geometrical length, as confirmed by the vibration displacement measurement of mirrors. These findings demonstrate the potential of the proposed system for application to primary acoustic standards and provide valuable insights toward achieving more accurate measurements.

The remainder of this paper is organized as follows. In Section II, the relationship between the optical frequency shift and the sound pressure shift is derived theoretically. Section III describes the experimental details, and the results are presented in Section IV. In Section V, the discrepancy in the sound pressure between measurements obtained with the proposed system and those obtained with a microphone is discussed. Finally, Section VI concludes this study.

## II. THEORY

This section describes the principle of sound pressure measurement through frequency shift detection using a Fabry-Pérot optical cavity.

### A. Relationship between the frequency shift and sound pressure

First, we consider a cavity with two mirrors with high reflectivity. The resonant frequency $\nu_q$ can be expressed as follows:

$$\nu_q = \frac{c}{2nL}(q + \varphi), \tag{1}$$

where the ratio of the spatial mode to the longitudinal mode of the cavity is expressed as

$$\varphi = \frac{1}{\pi}\cos^{-1}\left(\sqrt{\left(1-\frac{L}{R_1}\right)\left(1-\frac{L}{R_2}\right)}\right). \tag{2}$$

Here, $c$ denotes the speed of light, $n$ denotes the air's refractive index for light, $L$ denotes the geometrical distance between the mirrors, $q$ denotes the index of a longitudinal mode, and $R_1$ and $R_2$ are the radii of the curvature of each mirror, respectively [10]. We assume that only the lowest transverse electromagnetic (TEM) mode, $TEM_{00}$, is effectively coupled with the laser light.

The derivative of the resonant frequency with respect to pressure, $\partial v_q / \partial p$, is given by:

$$\frac{\partial v_q}{\partial p} = \frac{c}{2}(q+\varphi)\left(-\frac{1}{n^2 L}\frac{\partial n}{\partial p} - \frac{1}{nL^2}\frac{\partial L}{\partial p}\right) = -v_q\left(\frac{1}{n}\frac{\partial n}{\partial p} + \frac{1}{L}\frac{\partial L}{\partial p}\right), \tag{3}$$

provided that the mode number $q$ remains the same before and after the pressure fluctuates. In other words, the free spectral range (FSR), $\Delta v_{FSR} = v_{q+1} - v_q$, is sufficiently larger than the induced resonant frequency shift (FIG. 1). Although $\varphi$ varies slightly with changes in $L$, the sum $q + \varphi$ remains nearly constant because the contribution of $\varphi$ to $q$ is negligibly small — on the order of $10^{-6}$ when using 1555 nm laser. In practice, $q$ can be chosen to correspond to the laser frequency, and the exact value of $q + \varphi$ is unnecessary if the laser frequency is known.

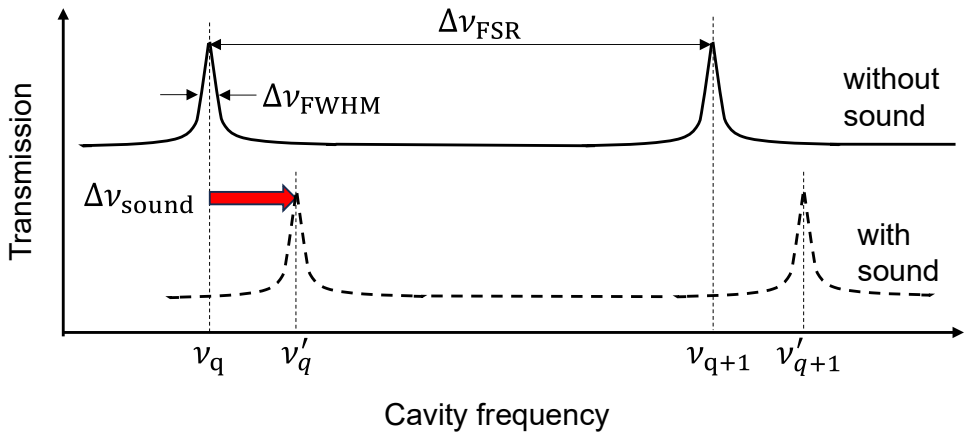

FIG. 1. Schematic illustration of the resonant frequency shift of the optical cavity.

According to Eq. (3), when the sound pressure $\Delta p$ is generated inside the cavity, the cavity frequency shift $\Delta \nu_{\text{sound}}$ is expressed as follows:

$$\Delta \nu_{\text{sound}} = -\nu_q \left( \frac{1}{n} \frac{\partial n}{\partial p} + \frac{1}{L} \frac{\partial L}{\partial p} \right) \Delta p. \tag{4}$$

Here, two factors contribute to the frequency shift. The first is $\frac{1}{n} \frac{\partial n}{\partial p}$, i.e., the change in the refractive index due to pressure, which we want to evaluate accurately. The second is $\frac{1}{L} \frac{\partial L}{\partial p}$, arising from mirror vibrations and cavity deformation, which we want to eliminate or isolate.

**B. Relationship between refractive index change and sound pressure**

In the following, the relationship between refractive index change and sound pressure is derived from the Lorentz–Lorenz equation and Poisson's law.

First, we consider the relationship between the refractive index and density. Beginning from the Lorentz–Lorenz equation for an ideal gas [11]:

$$\frac{n^2 - 1}{n^2 + 2} = A_R \rho_v, \tag{5}$$

we can derive:

$$n - 1 = \frac{n^2 + 2}{n + 1} A_R \rho_v, \tag{6}$$

where $A_R$ denotes the molar refractivity and $\rho_v$ denotes the molar density. In a laboratory environment, the refractive index $n$ of air is close to 1, and its change due to the pressure change is small. For example, using Ciddor's empirical formula, the change in $\frac{n^2+2}{n+1}$ is less than $10^{-6}$ when the pressure varies from 101325 Pa to 101335 Pa (other environmental conditions are listed in TABLE I). Accordingly, it can be treated as a constant $K$:

$$\frac{n^2 + 2}{n + 1} \approx K. \tag{7}$$

Thus, Eq. (6) can be rewritten using the constant $K'$ as follows:

$$\rho = K'(n - 1), \tag{8}$$

where

$$K' = \frac{M}{KA_R}. \tag{9}$$

Here, $M$ denotes the molar mass and $\rho$ denotes the density. Note that this relation is also known as the empirical Gladstone–Dale equation.

Second, the relationship between air density and sound pressure is discussed. Assuming an adiabatic process for pressure variation, Poisson's law gives:

$$\frac{P_1}{P_0} = \left(\frac{\rho_1}{\rho_0}\right)^\gamma, \tag{10}$$

where $P_0$ and $P_1$ are the pressures under static and sound pressure conditions, respectively, $\rho_0$ and $\rho_1$ are the corresponding densities, and $\gamma$ is the ratio of specific heats. Using Eq. (8), this becomes:

$$\frac{P_1}{P_0} = \left(\frac{K'(n_1 - 1)}{K'(n_0 - 1)}\right)^\gamma = \left(\frac{n_1 - 1}{n_0 - 1}\right)^\gamma. \tag{11}$$

By substituting $P_1 = P_0 + \Delta p$ and $n_1 = n_0 + \Delta n$, and assuming $\Delta n \ll n_0$, we obtain the following:

$$\frac{P_0 + \Delta p}{P_0} = \left(1 + \frac{\Delta n}{n_0 - 1}\right)^\gamma \approx 1 + \gamma \frac{\Delta n}{n_0 - 1}, \tag{12}$$

which leads to:

$$\frac{\Delta p}{\Delta n} = \frac{\gamma P_0}{n_0 - 1}. \tag{13}$$

Combining this with Eq. (4), the frequency shift caused by sound is given as follows:

$$\Delta \nu_{\text{sound}} = -\nu_q \left(\frac{1}{n_0} \cdot \frac{n_0 - 1}{\gamma P_0} + \frac{1}{L_0} \cdot \frac{\partial L}{\partial p}\right) \Delta p. \tag{14}$$

Assuming $\frac{1}{L_0} \cdot \frac{\partial L}{\partial p} = 0$ and using a 1555 nm laser, the frequency shift for a sound pressure of 1 Pa under static pressure of 101325 Pa is calculated as $3.604 \times 10^5$ Hz. The refractive index variation and the ratio of specific heats were calculated using Ciddor's equation and Cramer's equation, respectively, under the environmental conditions listed in TABLE I [12, 13].

TABLE I. Environmental conditions for calculation

|  | Symbol | Value | unit |
|---|---|---|---|
| Laser wavelength | $\lambda_{\text{laser}}$ | 1555 | nm |
| Static pressure | $P_0$ | 101325 | Pa |
| Temperature | $T_0$ | 23 | °C |
| $CO_2$ concentration | $x_{CO_2}$ | 430 | ppm |
| Relative humidity | $\phi_H$ | 45 | % |

Note that Eq. (5) does not hold under real-world conditions. When adjusted to real gas, it becomes the following:

$$\frac{n^2 - 1}{n^2 + 2} = \rho_v(A_R(T) + B_R(T)\rho_v + C_R\rho_v^2 + \dots), \tag{15}$$

where $B_R(T)$ and $C_R$ are the refractivity viral coefficients. In addition, $A_R(T)$ and $B_R(T)$ are temperature-dependent values. To estimate the uncertainty associated with using Eq. (5) instead of Eq. (15), we evaluated the relative contribution of $B_R(T)\rho_v + C_R\rho_v^2$ to $A_R$ using nitrogen ($N_2$) as an example, for which refractivity coefficients have been reported in the literature [14]. The calculated contribution was less than $10^{-5}$, which indicates that the uncertainty associated with this approximation is negligibly small.

### C. Uncertainty in sound pressure calculation from the frequency shift

This section briefly discusses the uncertainty involved in the aforementioned calculation. Here, we begin from the following model equation, which is a rearranged form of Eq. (14) but the dependency of the environmental conditions on $n_0$ and $\gamma$ are expressed explicitly:

$$\frac{\partial \nu_{\text{sound}}}{\partial p} = -\frac{c}{\lambda_{\text{laser}}} \left( \frac{1}{n_0(P_0, T_0, \lambda_{\text{laser}}, x_{\text{CO}_2}, \phi_{\text{H}})} \cdot \frac{n_0(P_0, T_0, \lambda_{\text{laser}}, x_{\text{CO}_2}, \phi_{\text{H}}) - 1}{\gamma(P_0, T_0, \phi_{\text{H}}) \cdot P_0} + \frac{1}{L_0} \cdot \frac{\partial L}{\partial p} \right). \quad (16)$$

The uncertainty budgets of $\frac{\partial \nu_{\text{sound}}}{\partial p}$ are summarized in TABLE II. Note that the uncertainties presented here are approximate estimates based on literature and empirical knowledge. In summary, the uncertainty in the laser wavelength was estimated with reference to the manufacturer's information, and the uncertainties related to the environmental conditions were estimated based on experience from our long-term monitoring of environmental conditions in the laboratory. In addition, the uncertainties associated with the empirical formulas were evaluated based on the literature [12, 13]. Refer to the footnotes of the TABLE II for further details. It should be noted that geometric length change was not considered in this evaluation.

As a result, the combined standard uncertainty associated with $\frac{\partial \nu_{\text{sound}}}{\partial p}$ was 0.1%. Considering that the uncertainty in the primary calibration of microphones using a reciprocity calibration system is 0.5%, the 0.1% value is not excessively large; however, in consideration of additional uncertainties, such as those from frequency fluctuation measurements and geometric length change, further reduction is desirable. The dominant contribution to the uncertainty is environmental condition fluctuations; thus, it would be feasible to reduce the uncertainty in $\frac{\partial \nu_{\text{sound}}}{\partial p}$ through real-time monitoring of environmental conditions and/or implementing a control system, especially a temperature control system.

TABLE II. Uncertainty budget of $\frac{\partial v_{sound}}{\partial p}$

| No. | Symbol | Uncertainty component | Value with estimated standard uncertainty | Sensitivity factor | Standard uncertainty (Hz/Pa) | Relative uncertainty in % |
|---|---|---|---|---|---|---|
| (1) | $\lambda_{laser}$ | Laser wavelength | $(1554.94 \pm 0.29)^a$ nm | $2.3 \times 10^{11}$ Hz/(Pa m) | $6.7 \times 10^1$ | $1.9 \times 10^{-2\ b}$ |
| (2) | $P_0$ | Static pressure | $(101325 \pm 870)^c$ Pa | $9.0 \times 10^{-4}$ Hz/(Pa²) | $7.8 \times 10^{-1}$ | $2.2 \times 10^{-4\ b}$ |
| (3) | $T_0$ | Temperature | $(23 \pm 0.29)^c$ °C | $1.2 \times 10^3$ Hz/(Pa °C) | $3.5 \times 10^2$ | $9.8 \times 10^{-2\ b}$ |
| (4) | $x_{CO_2}$ | $CO_2$ concentration | $(430 \pm 17)^c$ ppm | $1.9 \times 10^{-1}$ Hz/(Pa ppm) | $3.3 \times 10^0$ | $9.1 \times 10^{-4\ b}$ |
| (5) | $\phi_H$ | Relative humidity | $(45 \pm 2.9)^c$ % | $7.4 \times 10^0$ Hz/(Pa %) | $2.1 \times 10^1$ | $5.9 \times 10^{-3\ b}$ |
| (6) | $n_0 - 1$ | Ciddor's equation | – | – | – | $7.5 \times 10^{-3\ d}$ |
| (7) | $\gamma$ | Cramer's equation | – | – | – | $3.2 \times 10^{-2\ e}$ |
| (8) | $\frac{1}{L_0} \cdot \frac{\partial L}{\partial p}$ | Geometrical length change | To be determined | – | – | – |
| | | | Combined standard uncertainty of $\frac{\partial v_{sound}}{\partial p}$ | | | $1.1 \times 10^{-1}$ |

[a] The laser wavelength was assumed to be 1554.94 ± 0.5 nm based on the manufacturer's information (RIO, planex). A coverage factor of $1/\sqrt{3}$ was multiplied for the uncertainty budget, assuming the rectangular distribution.

[b] The relative uncertainty was calculated using $3.604 \times 10^5$ (Hz/Pa) (see Section IIB) as the reference.

[c] Environmental conditions with potential change were assumed to be $(101325 \pm 1500)$ Pa for static pressure, $(23 \pm 0.5)°C$ for temperature, $(430 \pm 30)$ ppm for $CO_2$ concentration, and $(45 \pm 5)\%$ for relative humidity. The experiment was conducted in an experimental room where the temperature and humidity were controlled. A coverage factor of $1/\sqrt{3}$ was multiplied for the uncertainty budget, assuming the rectangular distribution.

[d] The uncertainty in calculating $n_0 - 1$ was estimated based on the uncertainty in Ciddor's equation for calculating $n_0$.

[e] The uncertainty in Cramer's equation used to calculate $\gamma$ was reported as $3.2 \times 10^{-4}$.

## III. EXPERIMENT

### A. Optical sound pressure measurement

#### 1. *System*

FIG. 2 illustrates the developed optical sound pressure measurement system. The system comprises (a) an optical cavity, (b) an optical unit to lock the frequency of a narrow-linewidth laser to that of an optical cavity, and (c) a separate optical and measurement unit to read out the frequency fluctuations of the locked laser. These three units are described in detail in the following.

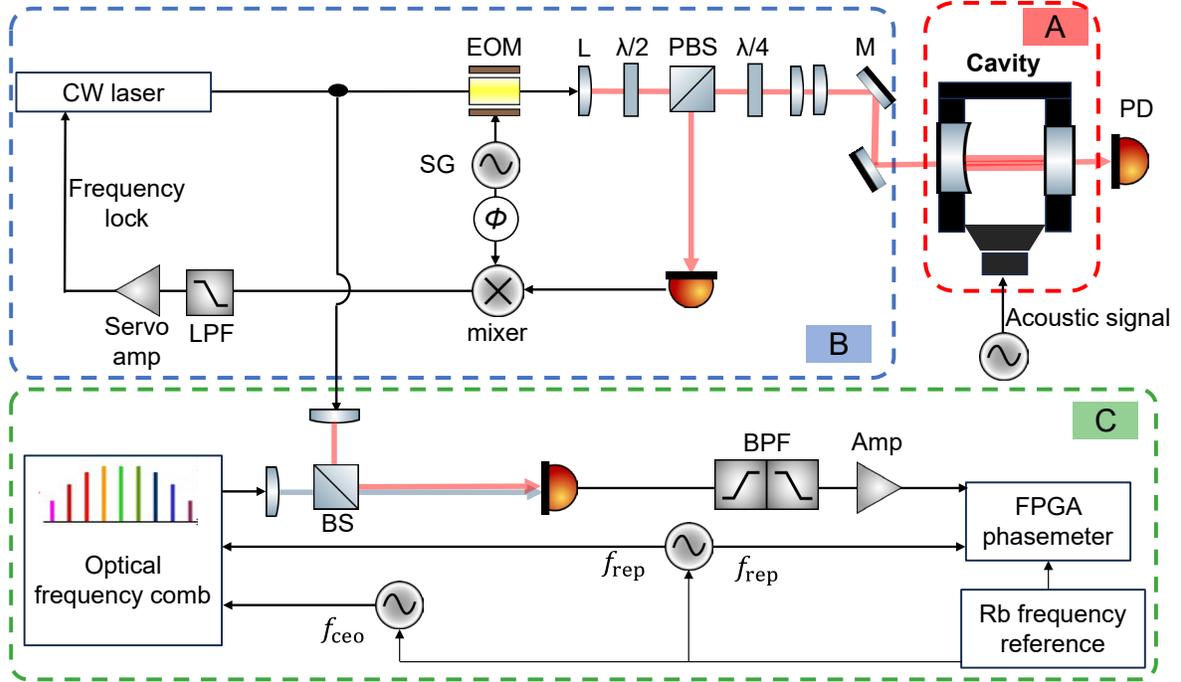

FIG. 2. Schematic of the optical sound pressure measurement system at NMIJ. EOM: electro-optic modulator; L: lens; M: mirror; $\lambda/2$: half-wave plate; PBS: polarizing beam splitter; $\lambda/4$: quarter-wave plate; PD: photo detector; SG: signal generator; $\phi$: phase shifter; LPF: low-pass filter; BS: beam splitter; BPF: bandpass filter; Amp: Amplifier.

a. *Optical cavity.* FIG. 3(a) shows a photograph of the constructed Fabry-Pérot optical cavity for sound pressure measurement. Here, two mirrors were mounted to a rectangular spacer with dimensions of 52 mm × 50 mm × 61 mm, to which a loudspeaker (KINGSTATE, KSSG3108) and a microphone were also attached. In addition, the platinum resistance thermometer was mounted on top of the spacer to monitor temperature change. To block the surrounding background noise and allow both the reference microphone and the optical system to sense the same sound pressure below a few kilohertz, a compact and sealed coupler with dimensions of several tens of millimeters was employed.

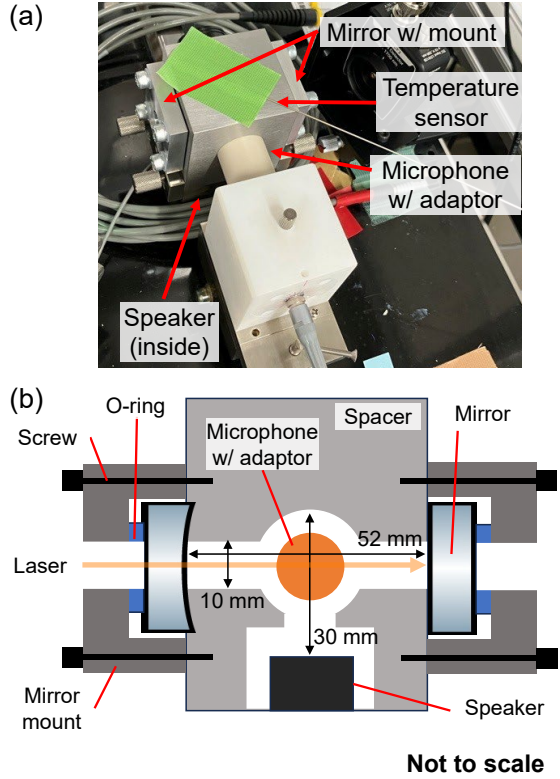

FIG. 3. (a) Photograph and (b) cross-sectional view of the constructed optical cavity.

The distance between the mirrors was designed to 52 mm, resulting in a calculated $\Delta\nu_{\mathrm{FSR}}$ value of 2.9 GHz. The nominal reflectance of both mirrors was specified as 99.97% by the manufacturer. Accordingly, the theoretical finesse $F$ and full width at half maximum (FWHM) of the cavity transmission peak $\Delta\nu_{\mathrm{FWHM}}$ were calculated to be 3000 and 0.97 MHz, respectively, using the following equations:

$$F = \frac{\pi\sqrt{r_1 r_2}}{1 - r_1 r_2}, \qquad (17)$$

$$\Delta\nu_{\mathrm{FWHM}} = \frac{\Delta\nu_{\mathrm{FSR}}}{F}, \qquad (18)$$

where $r_1$ and $r_2$ are the reflectance values of the two mirrors. In practice, the measured $\Delta\nu_{\mathrm{FWHM}}$ was 1.2 MHz, corresponding to a finesse value of approximately 2400. This discrepancy was probably due to the imperfect optical alignment and/or surface contamination of the mirrors.

As mentioned in Section II, performing accurate acoustic measurements requires the separation of the geometrical length changes from the refractive index changes induced by sound. For example, assume we want to measure the change in the refractive index caused by sound with an uncertainty of less than 0.1%. This corresponds to suppressing the contribution of $\frac{1}{L_0}\frac{\partial L}{\partial p}$ to 0.1% of $\frac{1}{n_0}\frac{\partial n}{\partial p}$, i.e., limiting $\frac{1}{L_0}\frac{\partial L}{\partial p}$ to $1.9 \times 10^{-12}$/Pa. In our optical cavity with length $L_0 = 0.052$ m, this requires either reducing the geometrical length change $\frac{\partial L}{\partial p}$ to less than $9.9 \times 10^{-14}$ m/Pa or accurately measuring the vibration displacement at this scale to subtract.

Therefore, we initially sought to minimize the geometrical path length change by devising the mirror mounting method and the choice of the spacer material, and then attempted to measure the vibration displacement accurately. The details of the vibration measurement are explained in Section IIIB. To suppress mirror motion in response to acoustic pressure, we designed a mirror-mounting structure in which the surface of the mirror is pressed against the surface of the cavity via an O-ring, as shown in FIG. 3(b). Since commercial mirror holders are typically designed for lateral support, a custom holder was fabricated for this configuration.

The use of materials with high-rigidity is effective in terms of suppressing the spacer deformation. In this study, stainless steel was selected rather than the ultra-low expansion (ULE) glass commonly used for cavity spacers, due to its higher Young's modulus and relative cost efficiency (TABLE III). If further improvement in rigidity is required, sapphire, which is used in coupler reciprocity method, or diamond, though more expensive, may be promising candidates.

A concern when using stainless steel is its relatively high thermal expansion coefficient compared with that of ULE glass, which may lead to cavity length fluctuations due to temperature changes. However, for the following reasons, we concluded that the length change associated with temperature fluctuations can be controlled adequately. First, although temperature variations in the spacer may be induced by environmental changes or heat generated from the loudspeaker and microphone, these variations are expected to occur in a frequency range below a few Hz. In contrast, the target frequency range in this study is above 100 Hz; thus, it is feasible to separate the effects of temperature drift using signal processing techniques. Furthermore, acoustic processes above 100 Hz can be regarded as adiabatic; therefore, no temperature variation is expected to occur directly as a result of the sound propagation.

TABLE III. Material candidates for optical cavity and their characteristics at room temperature

| Material | Mean coefficient of thermal expansion ($K^{-1}$) | Young's modulus (GPa) | Poisson ratio |
| --- | --- | --- | --- |
| Stainless[a] | $\sim 10^{-5}$ | 190 ~ 203 | 0.265 ~ 0.275 |
| ULE Glass[b] | $\sim 10^{-8}$ | 67.6 | 0.17 |
| Sapphire[c] | $\sim 10^{-6}$ | 340 ~ 370 | 0.28 ~ 0.33 |
| Diamond[d] | $\sim 10^{-6}$ | 1050 ~ 1210 | 0.18 ~ 0.22 |

[a] Reference[15]

[b] Reference[16]

[c] Reference[17]

[d] Reference[18]

b.  *Optical unit for locking frequency.*  The optical setup to lock the laser frequency to the resonance frequency of the optical cavity using the Pound-Drever-Hall technique is shown in the upper half of FIG. 2 [19]. Here, light from a narrow-linewidth laser (RIO, Planex; center wavelength: 1554.94 nm) was phase-modulated at 15 MHz by an electro-optic modulator (Exail, MPX-LN-0.1) and then incident on the optical cavity. The reflected light was detected using a photodetector and demodulated at the modulation frequency to produce an error signal, which was a voltage proportional to the frequency deviation between the laser and the cavity resonance. The error signal was fed back to the laser's drive current via a servo amplifier with a low-pass filter (Liquid Instruments, Moku:go). The cutoff frequency (−3 dB) of the low-pass filter was set to 15 Hz, and the unity gain frequency of the entire feedback circuit was measured to be approximately 30 kHz. This feedback enabled the laser frequency to track the resonance frequency of the optical cavity, which oscillated at the frequency of the acoustic signal. Note that the gain of the feedback circuit was considered in the calculation of the sound pressure. In practice, the laser control current was initially set to zero, and a small modulation current (a few mA) was applied to lock the laser frequency to the nearest resonant mode.

c.  *Frequency fluctuation measurement unit.*    As shown in the bottom half of FIG. 2, the frequency fluctuation of the laser locked to the optical cavity was measured using a phasemeter in reference to the frequency of an optical frequency comb [20]. The use of the comb offers a reference frequency with high precision and stability, and the phasemeter developed at NMIJ excels in measuring rapid and large-amplitude phase evolution. The repetition rate of the comb ($f_{\mathrm{rep}}$) was 61.7 MHz, and the carrier-envelope offset frequency ($f_{\mathrm{ceo}}$) was 21.4 MHz. Both $f_{\mathrm{rep}}$ and $f_{\mathrm{ceo}}$ were phase-locked to reference signals from signal generators. Beat signals were generated by the interference between the CW laser and the comb and detected using a photodetector. One of the beat signals was filtered with a tunable bandpass filter (passband width of 1.2 MHz), amplified, and then fed into the custom-made phasemeter based on a field-programmable gate array (FPGA phasemeter).

The FPGA phasemeter measured the phase difference between the beat note and a reference signal (in this study, the reference signal used for locking $f_{\text{rep}}$) using a proprietary algorithm [21]. Since the reference signal is sufficiently stable compared to the beat signal, the fluctuation in their phase difference was equivalent to the phase fluctuation of the beat signal. The frequency fluctuation of the beat signal was obtained by taking the time derivative of this phase difference variation.

### 2. Sound pressure measurement

Sounds with pressure levels of either 78 dB or 84 dB were generated by a loudspeaker inside the spacer across a frequency range of 100 Hz to 5 kHz at 1/3-octave intervals and measured using this system and a microphone (Brüel & Kjær, Type 4180, S/N: 2101405). A sine-wave voltage from the signal generator was injected into the loudspeaker. The generated sound pressure levels were set by considering the passband of the bandpass filter employed to extract one beat signal. The microphone had been calibrated in accordance with the acoustic standards of NMIJ with uncertainty of approximately 0.5%.

The optically derived sound pressure ($\Delta p_{\text{freq}}$) and sound pressure measured by a microphone ($\Delta p_{\text{mic}}$) were obtained as follows. The continuous waveforms of the frequency and microphone voltage were sampled using the FPGA phasemeter and a digitizer, respectively. The measurement time was adjusted to $n/f$ ($n$: an integer greater than or equal to 100, $f$: frequency of interest). Then, the sinusoidal amplitude at $f$ was extracted by applying the sine approximation method based on IEEE

Standard 1057:2017 [22]. This method estimates parameters of the sine wave, such as amplitude and phase, by performing a least-squares fit of continuously digitized waveforms to a sine wave at the desired frequency. The Hanning window was multiplied by each time waveform before the sine approximation [23]. Finally, $\Delta p_{\text{freq}}$ and $\Delta p_{\text{mic}}$ were obtained by multiplying the amplitude with the corresponding sensitivity. Measurements were repeated three times on different days to assess reproducibility.

### B. Vibration displacement measurement

FIG. 4 depicts the vibration displacement measurement system using heterodyne interferometers. The displacements of the two mirrors were measured synchronously using two heterodyne interferometers (Ono Sokki, LV-1800) placed face-to-face. Reflective tape was affixed slightly off-center on the outer surface of each mirror, and the displacement was measured by directing the laser beam onto the reflective area. The heterodyne beat signals were sampled and demodulated into displacement signals using the FPGA phasemeter.

The mirror vibrations consist of two components: one caused by the sound pressure pushing each mirror from inside the cavity, and another resulting from the overall motion of the cavity itself. In the current setup, the loudspeaker and the cavity are placed on the same base, which can transmit structural vibrations from the loudspeaker and cause the entire cavity to move. To eliminate the contribution of the overall vibration of the spacer, the signals from the two heterodyne interferometers were acquired synchronously, and the sum of the measured displacement waveforms was monitored.

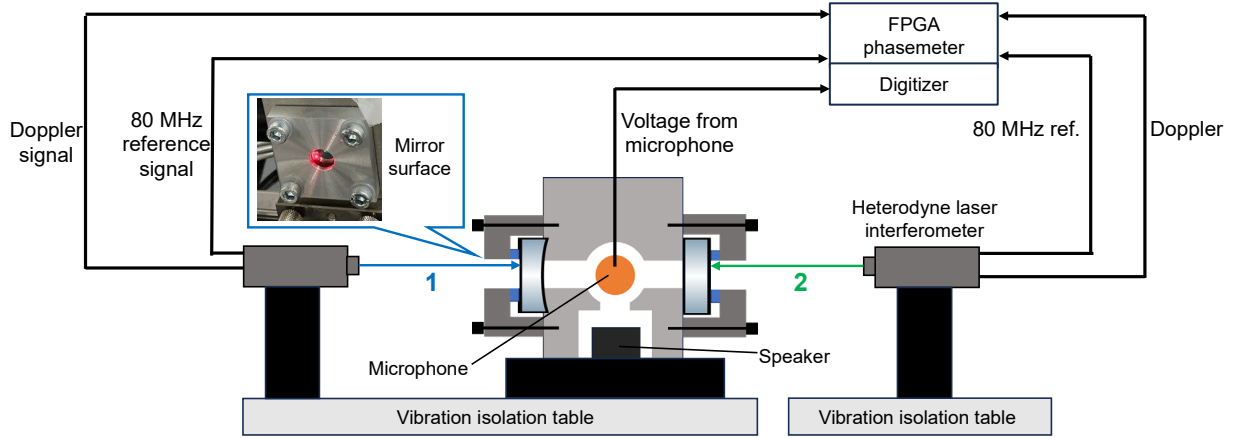

FIG. 4. Schematic of the vibration displacement measurement system. Due to positional constraints, one interferometer was installed on a vibration isolation table different from that used to support the cavity, whereas the other was placed on the same table.

Simultaneously, the acoustic pressure inside the spacer was monitored using a microphone. The output voltage waveform from the microphone and the summed displacement waveform were processed by sine approximation method, following the same procedure described in Section IIIA2. Then, the ratio of the summed displacement amplitude to the sound pressure amplitude calculated from the microphone voltage amplitude was calculated for each frequency.

As mentioned in Section IIIA1, the vibrational displacement of the mirrors was very small. To ensure a sufficient signal-to-noise ratio, the generated sound pressure level was set to approximately 100 dB (For frequencies above 2500 Hz where the input voltage exceeded the allowable limit, the sound pressure level was reduced to around 80 dB). Considering the high noise level of the interferometers in the low-frequency range, the measurements were performed above 640 Hz.

## C. Finite element analysis

Finite element analysis was conducted to evaluate the contribution of the cavity deformation to the change in geometrical length. The solid mechanics interface of COMSOL (COMSOL Multiphysics) was used. For this analysis, a simplified model of the cavity was constructed (FIG. 5). Specifically, the internal space of the spacer was modeled with only a cylindrical hole for the optical path, omitting the microphone housing section. In addition, both mirrors were assumed to be flat, and the contact surfaces between the mirrors and the spacer were defined using the penalty method.

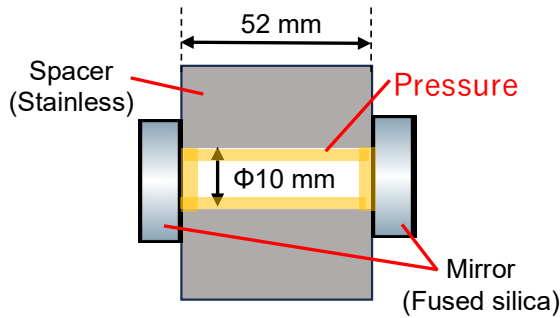

FIG. 5. Simplified cavity model for the finite element analysis.

## IV. RESULTS

### A. Results of sound pressure measurement

FIG. 6(a) shows a comparison of the $\Delta p_{\mathrm{freq}}$ and $\Delta p_{\mathrm{mic}}$ values. Note that we examine the ratio of $\Delta p_{\mathrm{freq}}$ to $\Delta p_{\mathrm{mic}}$ instead of comparing the sound pressure values directly. This is because the generated sound pressure lacked sufficient reproducibility; in other words, even when the same voltage is applied to the loudspeaker, the resulting sound pressure can vary by several percent depending on the measurement time and day. One plot of actual sound pressure data measured by the two methods is shown in Figure6(b). Below 1 kHz, the sound pressures measured by the two methods agreed within 5%; however, on average, $\Delta p_{\mathrm{freq}}$ was $(2.6\pm1.1)\%$ greater than $\Delta p_{\mathrm{mic}}$. This difference is attributed to the geometric length change between the mirrors—which was assumed to be zero—actually occurring and affecting the frequency fluctuation. The details and future measures are discussed in Section V. As for repeatability, the standard deviation in the $\Delta p_{\text{freq}}/\Delta p_{\text{mic}}$ ratio across three measurements was 2.5% at maximum. The results also demonstrated that higher sound pressure levels yielded better repeatability, which indicates that the signal-to-noise ratio (SNR) improvement directly contributes to repeatability.

In contrast, above 1 kHz, the difference between $\Delta p_{\text{freq}}$ and $\Delta p_{\text{mic}}$ became increasingly larger as the frequency increased. This discrepancy is probably due to the non-uniform sound pressure distributions in the spacer at higher frequencies, because the acoustic wavelength becomes comparable to the dimensions of the sound generation space. A peak was observed at approximately 4 kHz. Although detailed analysis using the finite element method is difficult due to the complex internal structure of the spacer, we primarily attribute this peak to acoustic resonance. The resonance may form a pressure antinode along the optical path, enhancing $\Delta p_{\text{freq}}$, while the microphone averages pressure over its membrane, thereby attenuating the observed $\Delta p_{\text{mic}}$ peak, as shown in Figure 6(b).

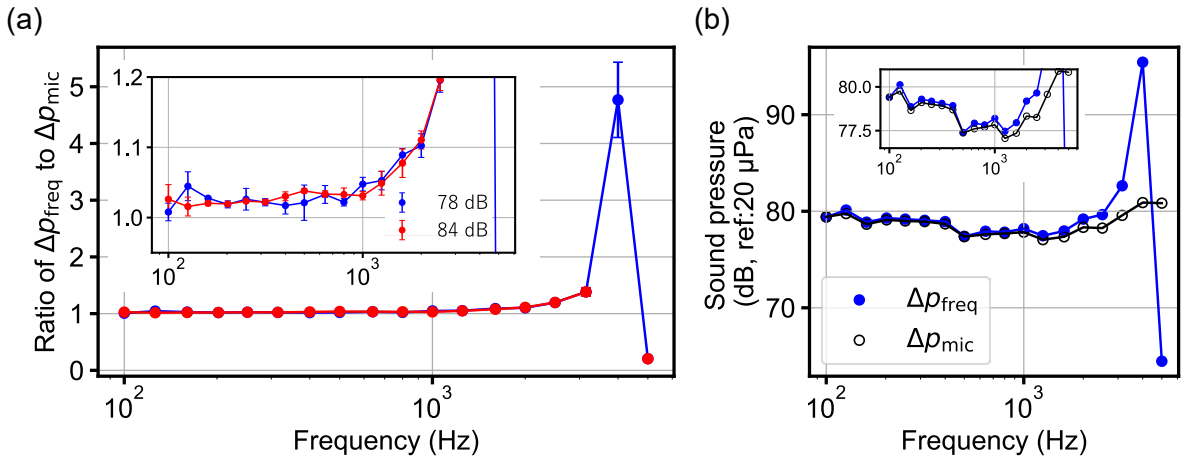

FIG. 6. (a) Frequency response of the ratio $\Delta p_{\text{freq}}$ to $\Delta p_{\text{mic}}$ when 78 dB (blue) and 84 dB (red) sound pressure were applied. (b) Comparison between $\Delta p_{\text{freq}}$ and $\Delta p_{\text{mic}}$ measured in a single test

with 78 dB sound pressure. The insets show an enlarged view. In (a), the error bar shows the standard deviation from three measurements. Note that the 4 kHz, 84 dB sound is not included because the frequency fluctuation induced by the sound was too large, causing the beat signal to fall outside the bandpass filter and making accurate measurement impossible.

To further verify whether the developed system can properly detect variations in resonance frequency caused by sound pressure, the frequency fluctuations of the beat signal were measured under static conditions and when a 1 kHz sound was applied. The time-domain fluctuation of the beat signal and its frequency spectrum are shown in FIG. 7(a) and FIG. 7(b), respectively. When the 1 kHz sound was applied, a frequency modulation was clearly observed at the corresponding acoustic frequency, confirming that the system can respond to sound excitation. Furthermore, as shown in FIG. 7(c) and FIG. 7(d), microphone measurements confirmed that the generated sound exhibited no harmonic distortion.

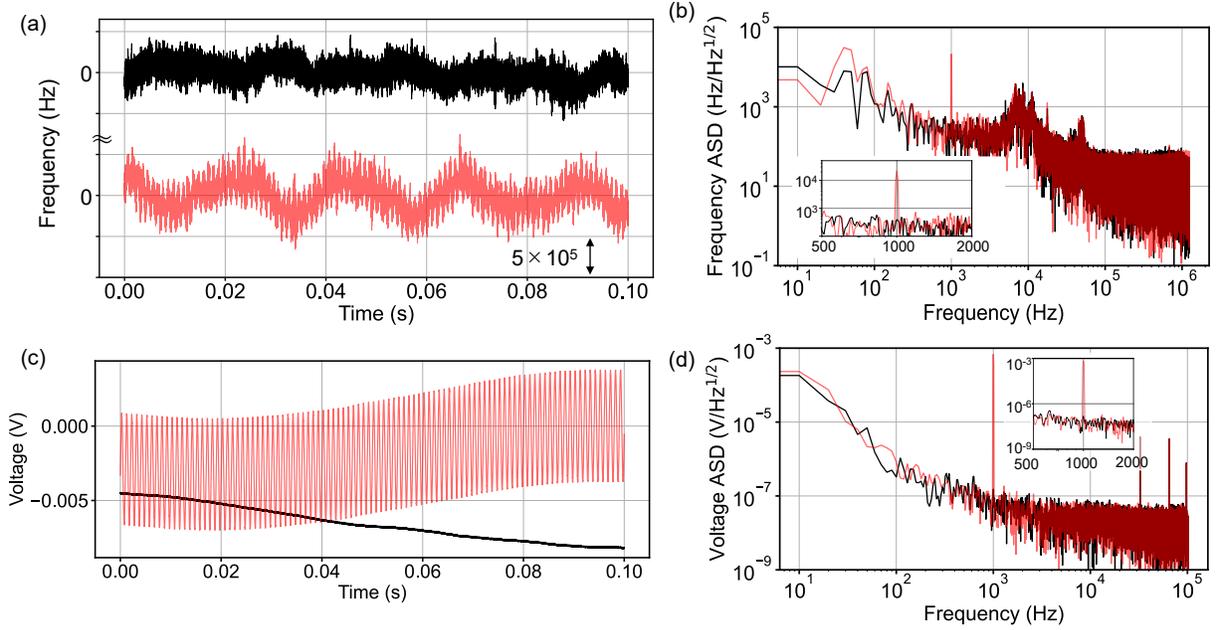

FIG. 7. Results of (a,b) beat note fluctuation measurement and (c,d) microphone voltage measurement without (black) and with 1 kHz 84 dB sound (red). (a,c) Time-domain and (b,d) frequency-domain. ASD: amplitude spectral density. In (a), the offsets are shifted for clarity. The insets in (b,d) show the enlarged view around 1 kHz.

### B. Results of vibration measurement

FIG. 8(a) shows that the ratio of the summed displacement to the sound pressure was nearly constant below 2 kHz, with an averaged value of $(1.82 \pm 0.04) \times 10^{-12}$ m/Pa. This corresponds to a 1.9% variation in the optical path length due to the refractive index change, which, although slightly smaller, agrees well with the 2.6% deviation reported in Section IVA. The difference between 1.9% and 2.6% may be attributed to the differences in the measurement conditions between the vibration displacement measurement and the optical sound measurement. For example, in the vibration displacement measurement, the generated sound pressure was approximately 10 times higher than that used in the optical sound measurement, and the displacement was measured at a point slightly off-center from the center of the mirror. The spectral analysis shown in FIG. 8(b) further confirms that the mirrors vibrated at the frequency corresponding to the generated sound, thereby supporting the validity of the measurement. FIG. 8(a) also indicates that the ratio gradually increased above 2 kHz and showed a peak at approximately 4 kHz, suggesting structural resonance may occur. However, the converted sound pressure from the summed mirror displacement was only about 10% of that from the microphone output, which cannot fully explain the several-fold difference between $\Delta p_{\text{freq}}$ and $\Delta p_{\text{mic}}$ in FIG. 6(a).

FIG. 8(c) compares the time-domain waveforms of Mirror1, Mirror2, and the summed waveform of Mirror1 and Mirror2. The raw waveforms were dominated by background noise near a few Hz and by 50 Hz power-line interference. After applying a fourth-order Butterworth filter centered at 1 kHz, it was observed that Mirror1 and Mirror2 oscillated in nearly opposite phase. This indicates that the mirror displacements included both the deformation induced by the sound pressure inside the spacer and the rigid-body motion of the entire spacer, and the rigid-body vibration component could be effectively subtracted using two synchronized interferometers.

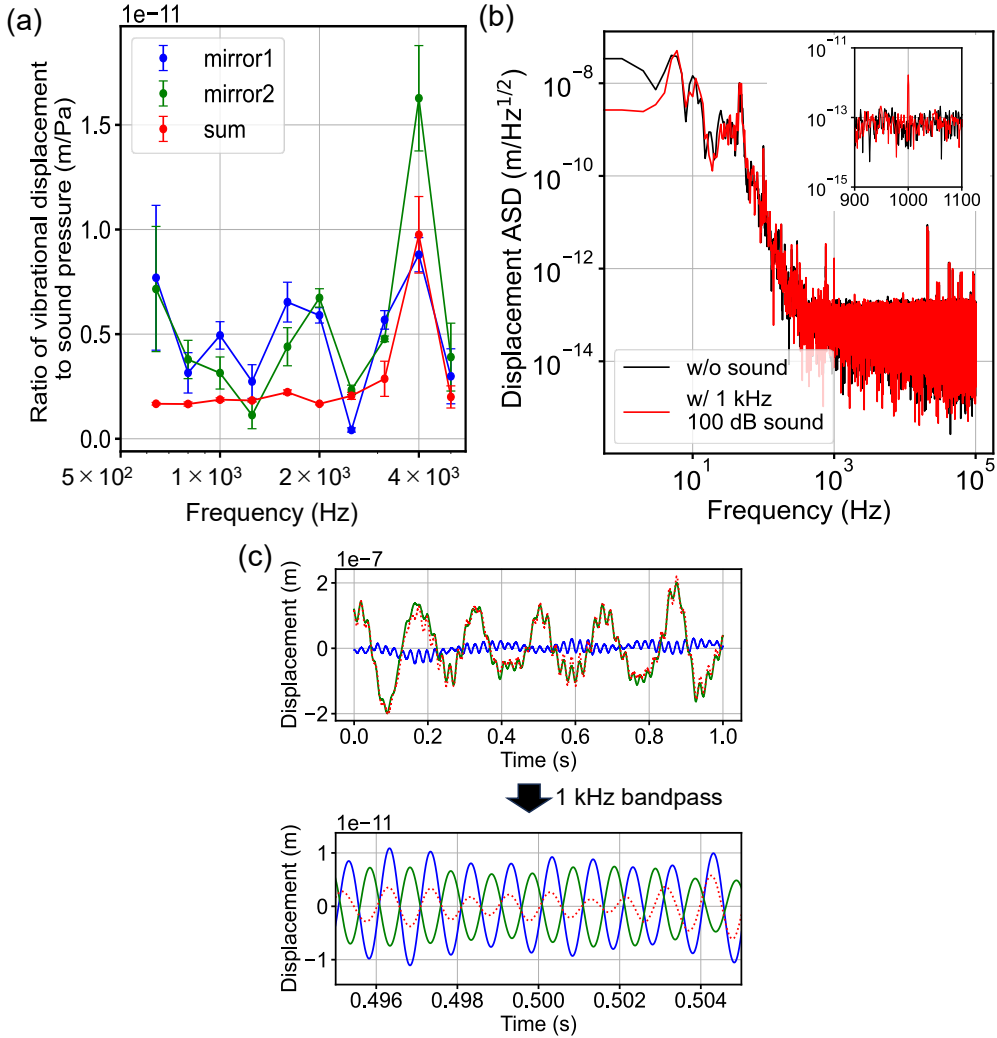

FIG. 8. Results of vibration displacement measurements. (a) Ratio of the displacement to sound pressure for Mirror1 (blue), Mirror2 (green), and the sum of the opposing mirrors (red). The error bars show the standard deviation from three measurements. (b) ASD spectra of summed displacement without (black) and with 1 kHz 100 dB sound (red). The inset in (b) shows the enlarged view around 1 kHz. (c) Time-domain displacement waveforms of Mirror1 (blue), Mirror2 (green), and the sum of the opposing mirrors (red) recorded during the generation of 1 kHz 100 dB sound. The upper panel shows the raw waveforms, and the lower panel shows the enlarged view of the waveforms after processing with a fourth-order Butterworth filter centered at 1 kHz.

## C. Results of finite element analysis

FIG. 9 shows the distribution of the deformation along the optical axis when a pressure of 0.15 Pa (corresponding to 78 dB sound) was applied. The deformation of the fused silica mirrors was more significant than that of the stainless steel spacer. The slope of the distance change between the mirror centers with respect to the pressure change was found to be $2.1 \times 10^{-13}$ m/Pa. This corresponds to 0.2% of the change in the optical length due to the change in the refractive index, which was about one tenth of the measured vibration displacement per unit pascal. The discrepancy arises because the FEA simulation and vibration measurements assess different components. The change in geometric length between the mirrors is considered to arise mainly from two contributions: (1) deformation caused by pressure applied to the spacer and mirrors, and (2) vibration-induced displacement resulting from insufficient fixation of the mirrors. Among these, the vibration measurements capture both contributions (1) and (2), whereas the FEA simulation accounts only for (1).

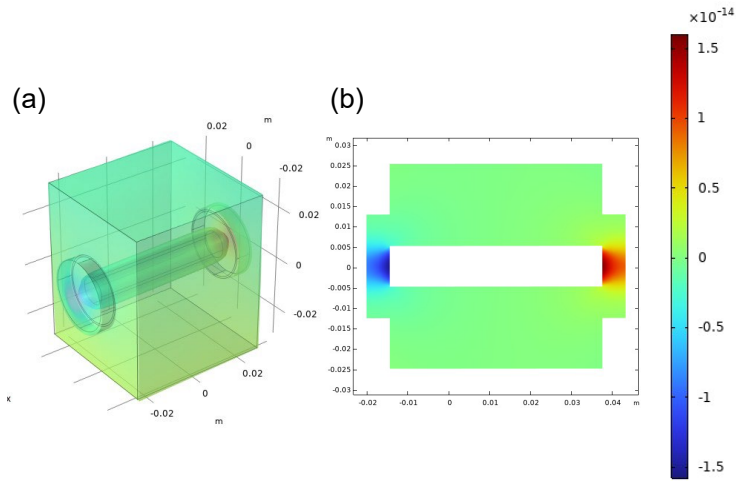

FIG. 9. Distribution of deformation along optical axis assuming a pressure of 0.15 Pa was applied to the internal surface. (a) Three-dimensional and (b) cross-sectional view.

## V. DISCUSSION

In this study, we developed an optical sound pressure measurement system based on frequency-fluctuation measurement using a Fabry–Pérot cavity. The sound pressures generated from 100 Hz to 5 kHz were measured using this system and a reference microphone. The results demonstrated that, although the sound pressures measured by the proposed system were larger than those measured by the microphone, the two methods agreed within a 5% deviation in the region from 100 Hz to 1 kHz.

Compared with previous work, the broader frequency coverage and higher precision achieved here can be attributed to improvements in the cavity structure and the frequency measurement system. First, we modified the mirror mounting configuration to suppress changes in geometric length and kept the cavity dimensions much smaller than the 1 kHz sound wavelength to ensure pressure uniformity within that frequency range. Second, by combining an optical frequency comb with a phasemeter, we enabled high-precision detection of frequency fluctuations.

To achieve our ultimate goal—the development of a direct sound measurement method that does not rely on microphones—two challenges must be resolved. First, it is necessary to demonstrate that the proposed system can measure sound pressure correctly. Because the "true" sound pressure value is unknowable, in practice, the sound pressure measured by a primary-calibrated microphone and that by this system are expected to agree within their uncertainties, and any remaining discrepancy requires a reasonable explanation. Second, the uncertainty of the proposed system must be smaller than that of a primary-calibrated microphone. In the following, we discuss the current progress and remaining challenges related to the first issue. The second issue will be addressed in future work.

Regarding the accuracy of the measured sound pressure, two types of deviation were observed between this system and the microphone. In the lower frequency region from 100 Hz to 1 kHz, the sound pressure obtained by this system exhibited a constant offset a few percent higher than the microphone measurements. In the higher frequency region from 1 kHz to 5 kHz, the deviation increased with frequency, reaching as much as a five-fold difference at 4 kHz. Three possible causes for these deviations are summarized as follows:

A. The optical sound pressure measurement system does not measure the sound pressure correctly.

B. The microphone does not measure the sound pressure correctly, which means that the microphone's primary-calibration sensitivity is incorrect.

C. The optical sound pressure measurement system and the microphone each measure different acoustic fields.

The offset in the lower frequency region can be attributed to cause A (i.e., an error on the optical measurement system side), whereas the higher frequency deviation can be attributed to cause C (i.e., the detection of different acoustic fields). Regarding cause B (incorrect microphone sensitivity), we cannot rule out this possibility because the only primary calibration method in this frequency range is the reciprocity method. However, previous international comparisons (CCAUV.A-K2) have shown that, between 20 Hz and 250 Hz, Laboratory Standard (LS) 1 microphone sensitivity calibrations by the National Physical Laboratory's laser pistonphone method and other participants' coupler reciprocity methods agreed within 0.5% [24], making it unlikely that the current few-percent deviations are due to cause B.

Concerning the offset deviation in the lower frequency region, the most likely explanation is that the change in geometric length between the mirrors was not fully suppressed. This hypothesis was supported by the vibration displacement measurement results, where the optical path length variation caused by mirror vibration was 1.9% of that caused by the refractive index change, which were in fair agreement with the offset deviation of 2.6%. In addition, the results of the finite element analysis demonstrated that the mirror displacement caused by the cavity deformation accounted for only about one-tenth of the observed mirror vibration displacement. This suggests that the primary cause of the change in geometrical length is the mirror vibration due to insufficient fixation rather than the deformation of the cavity itself. To suppress such mirror vibration in future implementations, the mirrors should be fixed more rigidly to the spacer—for example, by optical contact—instead of screw fastening with an O-ring. If further reduction is required, utilizing materials with high rigidity, e.g., sapphire, for both the spacer and the mirrors would be effective in terms of suppressing cavity deformation.

As for the higher frequency deviation, it is attributed that the optical measurement system and the microphone sample different acoustic fields. Although detailed analysis using the finite element method is difficult due to the complex internal structure of the current spacer, the maximum dimensions of the cavity space were approximately 30 mm vertically and 52 mm horizontally, which is close to the half-wavelength of a 4 kHz sound wave (approximately 43 mm). Therefore, at frequencies above a few kilohertz, acoustic resonances are expected to occur within the cavity, thereby making it difficult to maintain uniform sound pressure. To enable sound pressure measurements up to several kilohertz, we plan to miniaturize the coupler section to ensure sound pressure uniformity, and to simplify its geometry—for example, to a cylindrical shape—to facilitate numerical analysis. Note that fundamental difference remains between the optical and microphone measurements, i.e., the optical measurement system detects the line-integrated pressure change along a narrow optical path, whereas the microphone measures the spatial average pressure over the surface of its diaphragm. If the proposed system is applied for primary calibration in the frequency range above several kilohertz, particular attention should be paid to ensuring that both the optical system and the DUT sense the same sound pressure, regardless of the previously mentioned differences.

## VI. CONCLUSION

In this study, we demonstrated that combining careful optical cavity design with high-precision frequency measurement enables sound pressure measurements via frequency fluctuation measurements with deviations at the few-percent level in the 100 Hz to 1 kHz range. Several challenges remain in terms of improving accuracy and reducing uncertainty; however, we believe the findings of this study provide practical guidance for designing improved systems for optical sound measurements. Realizing direct sound measurement via optical techniques has the potential not only to free from dependence on specific microphones but also to expand the scope of acoustic measurement applications—because optical, non-contact methods can measure sound pressure even in environments (high temperature, high humidity, or intense sound pressure) where use of conventional microphones are impractical.

## ACKNOWLEDGMENTS

The authors acknowledge Dr. Yukimi Tanaka (NMIJ/AIST) for advice on finite element analysis, and Dr. Akiko Nishiyama (NMIJ/AIST) for guidance on optical alignment. This research was partially supported by the Suzuki Foundation (Fiscal Year 2025).

## AUTHOR DECLARATIONS

### Conflict of Interest

The authors have no conflicts to disclose.

## DATA AVAILABILITY

The data that support the findings of this study are available from the corresponding author upon reasonable request.

# CREDIT AUTHORSHIP CONTRIBUTION STATEMENT

**Koto Hirano**: Conceptualization, Methodology, Software, Data curation, Visualization, Investigation, Formal Analysis, Funding Acquisition, Writing - original draft. **Wataru Kokuyama**: Resources, Software, Investigation, Writing – Review & Editing. **Hajime Inaba**: Resources, Investigation, Writing – Review & Editing. **Sho Okubo**: Resources, Investigation, Writing – Review & Editing. **Tomofumi Shimoda**: Investigation, Software, Writing – Review & Editing. **Hironobu Takahashi**: Conceptualization, Methodology, Resources, Writing – Review & Editing. **Keisuke Yamada**: Conceptualization, Methodology, Resources, Writing – Review & Editing. **Hideaki Nozato**: Writing – Review & Editing, Funding Acquisition, Supervision.

## REFERENCES (NUMERICAL STYLE)